 \documentclass[prl,aps,twocolumn,showpacs]{revtex4}
\usepackage[dvips]{graphicx}
\usepackage{dcolumn}%
\usepackage{amsmath}%
\usepackage{amsfonts}%
\usepackage{amssymb}
\providecommand{\U}[1]{\protect\rule{.1in}{.1in}}
\newcommand{\be}{\begin{equation}}
\newcommand{\ee}{\end{equation}}
\newcommand{\bea}{\begin{eqnarray}}
\newcommand{\eea}{\end{eqnarray}}
\newcommand{\nn}{ \nonumber}

\begin{document}

\title{Modeling of a self-sustaining ignition  in a solid energetic material}
%% \title{Modeling of a self-sustaining ignition  in a solid energetic material}	
 %%\title{Self-sustaining ignition  in a solid energetic material: model and application}	

\author{Natalya A. Zimbovskaya}

\affiliation
{Department of Physics and Electronics, University of Puerto 
Rico-Humacao, CUH Station, Humacao, Puerto Rico 00791, USA}
%%\affiliation{Institute for Functional Nanomaterials, University of Puerto Rico, San Juan, Puerto Ruco 00931, USA}

\begin{abstract}
In the present work we analyze some necessary conditions for ignition of solid energetic materials by low velocity impact ignition mechanism. Basing on  reported results of {\it ab initio} computations we assume that the  energetic activation barriers for the primary endothermic dissociation in some energetic  materials may be locally lowered due to the effect of shear strain caused by the impact. We show that the ignition may be initiated in regions with the reduced activation barriers, even at moderately low exothermicity of the subsequent exothermic reactions thus suggesting that the above regions may serve as ``hot spots" for the ignition.  We apply our results to analyze initial steps of ignition in DADNE and TATB molecular crystals.
   \end{abstract}

%\pacs{72.15.Gd,71.18.+y}%%{71.18.+y, 71.20-b, 72.55+s}

\date{\today}
\maketitle

\section{i. introduction}

Solid energetic nitrocompounds often possess very desirable features for practical use, namely, the high performance (a measure of energy release) and the low sensitivity (a measure of an ease of the initiation of decomposition chemistry). Yet, detailed knowledge of chemical and physical processes characterizing ignition and detonation in these energetic materials (EM) is far behind their practical use. This lag in the understanding originates from an extraordinary complexity of the relevant physics and chemistry. Among the least understood are low  velocity impact ignition mechanisms in energetic solid nitrocompounds subjected to a dynamic mechanical loading. In particular, the nature of ``hot spots'' corresponding to these ignition mechanisms is not sufficiently clarified so far.

Suggested by Bowden and Youffe \cite{1}, a hot spot is supposed to be a small region in the interior of an inhomogeneous explosive solid where the ignition starts. These regions must have some special properties making them the preferable places for the ignition. These specific properties distinguish hot spots from the bulk of material. After the ignition happens in the hot spots, the reaction waves propagate from them to the main body of the explosive igniting the latter. A general agreement on the nature of hot spots is  not achieved yet. Nevertheless, it is reasonable to suggest that specific properties which indicate that a certain region is likely to become a hot spot, strongly depend on the ignition mechanism specifics. Several scenarios for the hot spots ignition have been explored over the years \cite{2,3,4,5,6,7,8,9,10,11,12,13,14,15}, and they are still being discussed.  
Presently discussed scenarios suggest that hot spots in solid energetic materials may  appear in regions characterized by enhanced dislocations density, porosity and grain-boundaries disorientation and/or at the interfaces between different  components in polymer-bonded explosives. The ignition process may be triggered by dynamic  mechanical loading (e.g. by shock waves) \cite{16,17,18} or by ultrasonic irradiation which may be effective in the case of composite materials \cite{19,20}. In some materials, hot spots may be generated by electromagnet stimulation \cite{21,22}, by laser beams \cite{22,23,24}, and by heating \cite{25,26}. In all probability, different mechanisms, singly or combined, control hot spots generation in different energetic solids under different conditions \cite{27,28}.
   The purpose of the present work is to contribute to understanding of the nature of hot spots in some solid nitrocompounds assuming that the ignition is  initiated by a low velocity impact. 

The  work is motivated by reported results of first-principle based computations of structural and electronic properties of several solid energetic materials including diamino-dinitroethylene $C_2H_4N_4 O_4 $ (DADNE) and triamino-trinitrobenzene $C_6H_6N_6O_6 $ (TATB) \cite{29,30,31,32}. Considered materials are layered molecular crystals. Their characteristics were analyzed assuming that they were submitted to the mechanical loading causing a shear strain in such a way that the adjacent layers were shifted with respect to each other. The obtained results  suggest that at the certain directions of the external stress the shear-strain displacements of the adjacent layers in DADNE molecular crystals may bring local variations of  important electronic characteristics. 
In particular, the band gap reduction and  a significant local lowering of the activation barrier for the detachment of $ NO_2 $ groups were predicted for DADNE molecular crystals \cite{29}. It was suggested that 
 the band gap reduction  can trigger sensitivity of DADNE crystals to optical and/or thermal stimuli via electronic excitations \cite{30}, which can lead to either direct bond breaking or high population of vibronically hot states \cite{27}. Also, it was  conjectured that the mechanical relaxation of the system may cause local heating of the material \cite{32}. 
All these changes were expected to be manifested in small regions  characterized by the increased density of defects appearing as a result the shear-strain deformation. They should appear within a short time after the strain is generated by, e.g. an impact or a shock wave. 
  The described results give grounds to explore a possible part which the deformed regions may take on in the process of ignition originating from the mechanical low velocity impact. Here, we carry out the corresponding analysis using chemical kinetic and heat balance equations.  

Usually, decomposition of solid EM  starts from a primary endothermic reaction, which
 is followed by a sequence of exothermic reactions.  Obviously, for ignition to develop  the net energy released in the result of the complete sequence of exothermic reactions should  substantially  exceed the amount of energy needed to support the primary endothermic step started by the external perturbation. However, the initiation chemistry in energetic compounds is a complex process, and the involved exothermic reactions are hardly expected to simultaneously develop. Therefore, it may happen that exothermicities of the reactions immediately following  the first endothermic step of the decomposition process are rather low. At the same time, the external disturbance may be too short-lived to affect a sufficient number of molecules in the explosive. This could occur due to the small size of hot spots and for some other reasons. In such cases, further development of the chemical decomposition crucially depends on the relation between the amount of energy consumed in the course of the primary endothermic reaction and the one released in the course of early exothermic reactions. When these reactions do not give out enough energy to  sustain the primary endothermic reaction, the whole ignition chemistry becomes quenched at the very beginning.  

In the second Section of this work we put forward a kinetic model to show that reduction of the activation barrier for the primary endothermic reaction
 could provide favorable conditions for initiated chemical reactions to be effectively transferred into a self-supported regime.
 We use this model to analyze the heat balance due to the primary endothermic and immediately subsequent exothermic chemical reactions occurring at the earliest decomposition stage. We show that the self-supported ignition regime and significant heating of the material become possible within the regions where 
the activation barriers for the primary reaction are reduced whereas the decomposition chemistry still rapidly decays beyond these regions. In the next Section we apply the obtained results to show that the small deformed regions with the reduced activation barriers for endothermic detachment of $ NO_2 $ groups may serve as hot spots in the ignition of DADNE molecular crystals.  These regions could appear as a result of mechanical impact or shock-wave propagation.

\section{ii. Model and results}

In the following analysis we assume  that the considered EM includes some special regions which are mainly distinguished from the bulk of material  by the locally reduced activation energy for the primary endothermic reaction and by the original temperature $ T_0 $ in the interior of these regions which may differ from the ambient temperature $ T_a. $ To simplify further calculations we reduce the complex chemical kinetics to a sequence of two first order reactions (one being endothermic and another one exothermic) with the reaction rates $ w_1 $ and $w_2, $ respectively. The corresponding kinetic equations may be written in the form \cite{33}:
  \be
  - \frac{dn_1}{dt} = w_1n_1 - D_1 \nabla^2 n_1 ,  \label{1}
  \ee
  \be
 - \frac{dn_2}{dt} = w_2n_2 - w_1 n_1 + D_1 \nabla^2 n_1 - D_2 \nabla^2 n_2.           \label{2}
  \ee
  When the secondary reaction is of the second order, the equation (\ref{2}) must be replaced by:
   \be
- \frac{dn_2}{dt} = w_2 n_2^2 - w_1n_1 + D_1 \nabla^2 n_1 - D_2 \nabla^2 n_2 .
                 \label{3}             \ee
 Here, $n_{1,2} $ are the reactants  concentrations expressed in $cm^{-3}. $ Within the chosen model, the reversion of $ n_2 $ to $ n_1 $ is taken to be unimportant, which is true for the ignition process. The reaction rates are supposed to depend on temperature following Arrhenius behavior:
  \be
 w_{1,2} (T) = A_{1,2} \exp \left(-\frac{E_{1,2}}{kT}\right)    \label{4}
  \ee
 where $ A_{1,2} $ and $E_{1,2} $ are the frequency factors and activation energies for the considered reactions, and $ k $ is the Boltzmann constant.  The terms including Laplacians of the reactant concentrations $(\nabla^2n_1,\ \nabla^2n_2) $ describe diffusion of the reactants to/from the considered region $D_{1,2} $ being the diffusion coefficients. We describe the time dependence of temperature in this special region  by means of the heat balance equation:
  \be
 \frac{dT}{dt} = \frac{Q}{\rho C} \left(-\frac{dn_2}{dt} \right) + \frac{N_a E_1}{\rho C} \left (\frac{dn_1}{dt} \right) - \frac{T - T_a}{\tau}  .   \label{5} 
    \ee 
  In this equation, the first two terms correspond to the contributions coming from the chemical reactions, and the last term originates from the heat transfer between the considered region and the surrounding material. Here, $Q $ is the exothermicity of the secondary reaction per one mole of the material, $ \rho, C $ are the material density expressed in $mol/cm^3 $ and the molar heat capacity, $ N_a $ is the Avogadro number,  and the parameter $ \tau $ is the characteristic time of the heat transfer. In the heat balance equation (\ref{5}), the energy consumed in a single act of the primary reaction is set equal to the corresponding activation energy $ E_1. $ This is a reasonable approximation for some reactions such as unimolecular bond dissociation. The chemical reactions of this kind may serve as primary steps in the decomposition of energetic nitrocompounds as discussed in the next section. This gives the necessary justification for the employed approximation. Usually, characteristic rates of chemical reactions involved in the ignition processes in energetic materials significantly exceed characteristic speeds of mass diffusion, therefore we omit diffusion terms from the kinetic equations (\ref{1})--(\ref{3}). Also, in writing out the heat balance equation for our system we assume that the temperature is spatially uniform over the considered region, which accounts for the absence of a term proportional to the temperature gradient in the Eq. (\ref{5}).

Using the dimensionless temperature $ \theta $ and time $ \xi :$ 
  \begin{align}
&\quad \theta = \frac{E_2}{kT_0} \frac{T - T_0}{T_0} ; 
  \nn\\
& \xi = A_2 \exp \left(-\frac{E_2}{kT_0}\right) t;                  \label{6}
  \end{align}
 and introducing dimensionless parameters:
\begin{align}
  \mu_{1,2} = \frac{kT_0}{E_{1,2}};   &\qquad  \qquad  \epsilon = \frac{k\rho CT_0^2}{E_2 Q N} ; 
   \nn \\
 \delta = \frac{\epsilon \exp(1/\mu_2)}{A_2 \tau} ; &\qquad \qquad
  \alpha = \frac{N_a E_1}{Q};
   \nn\\
   a= {A_1}/{A_2};        \label{7}
   \end{align}
  we may rewrite Eqs. (\ref{1})-- (\ref{3}) and (\ref{5}) in the following form:
  \be
 - \frac{dn_1}{d\xi} = an_1 \exp(\lambda_1);     \label{8}
 \ee
 \be 
  -\frac{dn_2}{d\xi} = n_2 \exp (\lambda_2) - an_1 \exp(\lambda_1);    \label{9}
  \ee 
    \be 
  -\frac{dn_2}{d\xi} = n_2^2 \exp (\lambda_2) - an_1 \exp(\lambda_1);    \label{10}
  \ee
  \be
\epsilon \frac{d\theta}{d\xi} = \frac{1}{N} \left(-\frac{dn_2}{d\xi}\right) + \frac{\alpha}{N}  \frac{dn_1}{d\xi} - \delta(\theta + \theta_a)                      \label{11}
  \ee
 where   %%$ \lambda_{1,2} $ are functions of temperature, namely:
   \be
 \theta_a = \frac{E_2}{kT_0} \frac{T_0 - T_a}{T_0}  ,  \label{12} 
   \ee
  $\lambda_i \ (i=1,2) $ are functions of temperature, given by:
   \begin{align}
&\lambda_1(\theta) = \frac{\mu_1 - \mu_2 + \mu_1\mu_2\theta}{\mu_1\mu_2(1+\mu_2\theta)}, 
   \nn\\ 
&\qquad  \lambda_2(\theta) = \frac{\theta}{1 + \mu_2\theta},               \label{13}
   \end{align}
 and $ N $ is the initial reactant concentration for the primary reaction. Omitting this reaction from consideration, we can reduce our system  to the pair of equations employed in the earlier work \cite{34} to describe thermal explosions.

Now, we numerically solve Eqs. (\ref{8}),(\ref{9}),(\ref{11}) using the initial conditions
   \begin{align}
 n_1(0)  & = N,  \nn \\  n_2(0)  & = 0, \nn\\ \theta(0) & = 0.         \label{14}
  \end{align}
 First, we consider the solution of our system within a localized region where the activation barrier for the endothermic reaction is lowered, which reduces  the difference between activation energies $E_1 $ and $ E_2.$ Correspondingly, we put $ E_1 = 1.5 eV $ and $ E_2 = 1 eV. $ Also, we assume that the chemical reactions within the region start at the temperature $ T_0 = 500 K.$ We emphasize that this high temperature is  characterizing only the considered special region. The ambient temperature $ T_a $  may be significantly lower than $T_0. $  To estimate  the parameter $\delta$ describing the intensity of heat transfer we accept the following physically meaningful values for the relevant parameters: $ \rho = 10^{-2} mol/cm^3,\ C =200 J/mol.K,\ A_2 \approx 10^{16} 1/s. $ The characteristic time of heat transfer $ \tau $ could be estimated basing on the results for the characteristic times for the phonon-phonon and phonon-vibron relaxation in shock-compressed materials \cite{35}, which gives $ \tau \sim 10^{-11} -10^{-9} s. $ Also, we may estimate the relaxation time using the experimental data on the heat capacities and thermoconductivities $(\kappa) $ of solid energetic nitrocompounds given by Tarver {\it et al} \cite{27}. For instance, choosing thermal properties of a typical nitrocompound TATB, namely: $ C = 0.4 cal/g\cdot K,\ \kappa = 5.5 \times 10^{-4}cal/(cm\cdot s\cdot K) $ and assuming that the volume of special region takes on  values of the order of $ 10^3 nm^3$ we get $ \tau\sim 10^{-9} s, $ which does not contradict the results  of phonon-based analysis.
Using these estimations we get $ \delta \sim 0.1 \div 10. $ 

  Time dependences of the concentrations $ n_1 $ and $ n_2 $ are presented in the Figs. 1,2. As expected, the concentrations relationship crucially depends on the ratio of the reactions rates. Using the assumed values for the activation energies we may estimate this ratio at temperatures close to the initial temperature $ T_0 $  as follows:
   \be 
 \frac{w_1}{w_2} \approx a \exp\frac{E_2 - E_1}{kT_0}  \sim a\times 10^{-3}.     \label{15}
  \ee
 When the parameter $ a $ takes on values smaller than $ 10^3 ,\ w_1$ is lower, than $ w_2 $, and the concentration $ n_1 $ exceeds $ n_2 $ all the time, as shown in the Fig. 1. However, when the ratio $ A_1/A_2 = a $ is  of the same order or greater than $ 10^3 ,$ the concentration $ n_2 $ surpasses $ n_1 $ shortly after the beginning of the reactions (see Fig. 2). 
 The relationship between the reactants concentrations is not very important when the exothermicity of the secondary reaction significantly exceeds the  energy consumed at the endothermic step. However, in the case of moderate or low exothermicity this relationship becomes crucial. Only sufficiently high $n_2 $ (exceeding $n_1)$ may provide an adequate energy release for sustaining the primary endothermic  reaction thus supporting the whole chemistry eventually leading to the explosion. In further calculations we put $ a= 10^4. $

\begin{figure}[t]  %%%fig. 1
\begin{center}
\includegraphics[width=8.8cm,height=4.5cm]{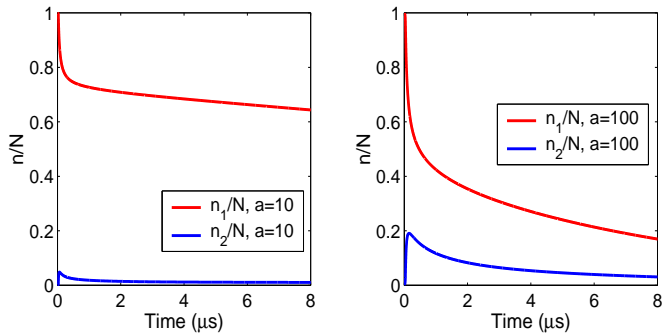}
\caption{ Time dependences of the reactant concentrations for the primary  and secondary reactions  under the condition $w_1(T_0)< w_2(T_0).$ The curves are plotted using Eqs. (\ref{8}),(\ref{9}),(\ref{11})  at $E_1 = 1.5eV,\ E_2 = 1eV,\ T_0 = 500K,\ A_2 = 10^{16} s^{-1},\ \alpha =0.4, \delta = 0.5 . $    %%%,\ a=10 $ (left panel) and $ a=100$ (right panel).
}  
\label{rateI}
\end{center}
\end{figure}

The direct evidence of the self-sustaining reaction  is the temperature rise. Obviously, the temperature dependences on time and reagents concentrations are determined by the relation of the value of the decomposition energy for the primary endothermic reaction and the exothermicity of the secondary reaction, and by the intensity of heat transfer between the separated region and its surroundings.  
At certain value of the parameter $ \delta, $ the effect of heat transfer strongly depends on the difference between the temperature in the interior of the separated region  $ T_0 $ and the ambient temperature $T_a. $
   Now, we analyze the temperature dependences of time and concentration of the primary reactant $ n_1 $ for various values of the parameter $ \alpha ,$ at two different ambient temperatures assuming the remaining parameters of the model to be fixed. The results shown in the  Fig. 3 give grounds to conjecture that the self-sustaining reaction may occur even at $\alpha = 1.2$ on the condition of a weak heat transfer provided by the closeness of the temperatures $ T_0 $ and $ T_a. $ At high exothermicity of the secondary reaction $ (\alpha =0.2)$ we observe a smooth but speedy temperature rise up to $ 650 K $ with the subsequent sharply defined fall down to the value close to the ambient temperature $ T_a. $ This conjecture is further supported by the plottings presented in the right panel of the figure. When the concentration $ n_1 $ becomes much smaller than the initial concentration $ N,$ we observe the speedy temperature rise for $ \alpha $ as large as $1.2. $ Such behavior is inherent to chemical reactions occuring within the intermediate regime between the slow reaction development and explosion \cite{34}. 
Here, we remark  that the temperature fall succeeding its previous rise appears in consequence of the oversimplicity of the adopted model where only two chemical reactions are accounted for. As was mentioned above, the decomposition chemistry includes a cascade of reactions and the energy release at the steps following the first two considered within the model should prevent the temperature fall shown in the Fig. 3. Nevertheless, the model predicts the temperature rise in the beginning  of the decomposition chemistry. This indicates the self-sustaining first phase of decomposition process which may occur within the special regions despite  the low exothermicity of  the first exothermic step immediately following after the primary endothermic reaction. 

\begin{figure}[t]  %%%fig. 2
\begin{center}
\includegraphics[width=8.8cm,height=4.5cm]{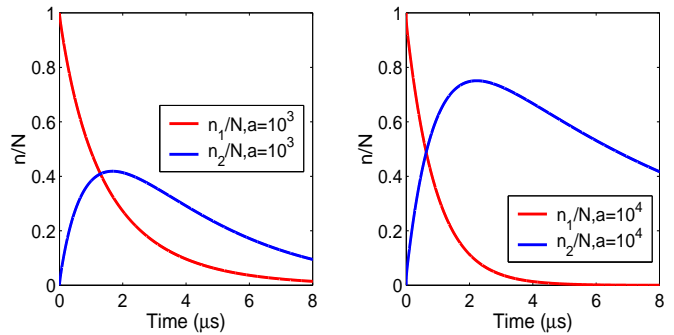}
\caption{ Time dependences of the reactant concentrations for the primary and secondary  reactions   plotted for the case when $w_1(T_0) > w_2(T_0). $  %% In plotting the curves we assumed that $ a=10^3 $ (left panel) and $ a= 10^4$ (right panel). 
 The remaining parameters are the same as in the Fig. 1.
}  
\label{rateI}
\end{center}
\end{figure}

 When $T_a $ is significantly lower than $ T_0 , $ the strengthening of the energy transfer between the separated region and the surrounding bulk material causes more stringent conditions for self-sustaining chemistry to occur inside the region. As shown in the left panel of the Fig. 4, high exothermicity of the secondary reaction $(\alpha= 0.2)$ is required to provide a slight temperature rise in the interior of such a region. We emphasize that the 
crucial condition for the self-sustaining chemistry to occur is the excess of the primary endothermic reaction rate over that of the subsequent exothermic reaction. When this happens we may observe a significant accumulation of the second reactant, and soon after the initiation$n_2 $ becomes greater than $ n_1. $ This enlarges the possible number of single acts of the exothermic reaction and, in consequence, the possible amount of the released energy. Under the described condition the separated regions may take on the part of hot spots where the decomposition chemistry starts.

Now, we turn our attention  to the main body of the energetic material. Assuming that $ E_1 \approx 2eV, $ and $ T_a=T_0 = 500 K $ we compare the reaction rates for the primary and secondary reactions, and we get $ w_1(T_a)/w_2(T_a) \approx 3a\times 10^{-6}.$ At lower temperatures $T_a $ the ratio or reaction rates takes on even smaller values. This means that the ratio $ A_1/A_2$ must be of the order of $ 10^7$ or greater to provide accumulation of the secondary reactant. So, even at $ T_a $ close to $ 500 K $ the concentration versus time curves in the bulk of material will be similar to those shown in  the  Fig. 1 except for the case of extremely fast primary reaction. The most important feature of such curves is the low concentration $ n_2 $ which always remains much smaller than $ n_1. $  Therefore, we may not expect a sufficient amount of energy to be released as a result of the secondary reaction to support the primary endothermic reaction  without the energy supplied by the external perturbation. This conclusion is illustrated by the right panel of the Fig. 4. The temperature versus time curves plotted in the figure demonstrate the fall of temperature, which indicates the impossibility of the self-sustained chemistry. This indicates the lower sensitivity of the bulk of material compared to that of separated regions. %%To start decomposition chemistry in the main body of EM one must significantly rise the temperature. For instance, using the assumed values for the activation energies and the frequency rates $ (a = 10^4),$ we may show that accumulation of the secondary reactant could occur at $ T_0 \approx 2000 K. $

\begin{figure}[t]  %%%fig. 3
\begin{center}
\includegraphics[width=8.8cm,height=4.5cm]{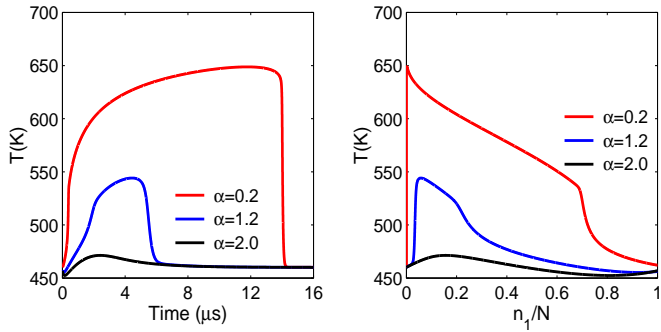}
\caption{ Temperature dependences on time (left panel) and concentration of the primary reactant (right panel) in the interior of the separated region computed at $E_1 = 1.5 eV,\ E_2 = 1 eV,\ T_0 = 500 K,\ T_a = 450 K,\ \delta = 0.1,\ a =10^4 .$  %%for $\alpha =0.2 $ (dash-dotted lines), $1.2 $ (dashed lines), $2$ (solid lines). 
}  
\label{rateI}
\end{center}
\end{figure}

Under normal conditions, practical EM do not demonstrate spotaneous combustion and/or detonation. In these materials chemistry leading to explosion is initiated by an external factor. %%%Initiation mechanisms provided by various external factors vary corresponding to the physical nature of these factors. Here, we concentrate on explosion initiating factors which could directly start the initial endothermic reaction by giving to the original molecules  energy necessary to break the relevant bonds. A laser beam is an obvious example of such a factor.
 While the external factor acts upon the material, certain amount of the primary reactant is consumed which is accompanied by the production of the corresponding amount of the secondary reactant.

To include this effect into consideration, we modify our model. We replace the heat balance equation (\ref{11}) by the following equation:
   \be
 \epsilon \frac{d\theta}{d \xi} = \frac{1}{N} \left(-\frac{dn_2}{d\xi} \right) + \frac{f(p)\alpha}{N} \frac{dn_1}{d\xi} - \delta (\theta + \theta_A).   \label{16}            
  \ee
 Here, $p = \Delta n_1/ N $ is the portion of the primary reactant consumed due to the external factor action, and $ f(p) $  is a monotonous function, which varies between 1 and 0 while $ p $ varies between 0 and 1. So, at $p=0 $, our heat balance equation is reduced to the previous form (\ref{11}). Within the unrealistic limit $ p=1 $ (when the initial reactant is completely consumed as a result of the action of the external factor), the second term describing contribution from the primary endothermic reaction disappears from the heat balance equation. The form of the function $ f(p) $ must be separately explored, and it may vary for different materials. Here, we choose the simplest possible approximation, namely: $f(p) = 1 - p. $ Also, we assume that $ t=0 $ corresponds to the instant when the external factor ceases to act, therefore initial conditions for the reactant concentrations must be replaced by:
  \be
 n_1 (0) = N (1 - p ); \qquad n_2(0) = \tilde N          \label{17}
   \ee
 where $ \tilde N $ is a nonzero initial concentration of the secondary reactant.

\begin{figure}[t]  %%%fig. 4
\begin{center}
\includegraphics[width=8.8cm,height=4.5cm]{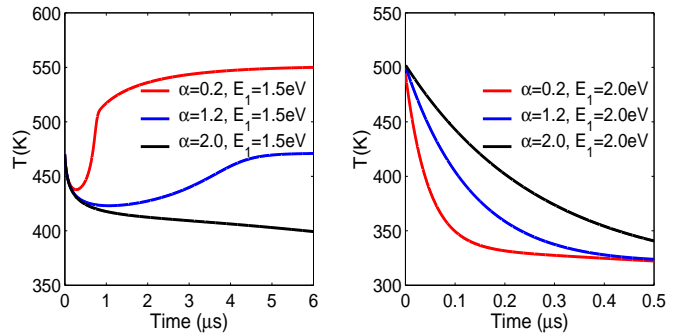}
\caption{ Temperature dependences on time in the interior of the separated region (left panel) and outside the latter (right panel). The curves are plotted for $T_a = 300 K,\ T_0 = 500 K,\ E_2 = 1eV,\ \delta = 0.1,\ a= 10^4.$
 %% assuming $ E_1 = 1.5eV $ (left panel) and $ E_1 = 2eV $ (right panel). 
}  
\label{rateI}
\end{center}
\end{figure}

Temperature dependences on time within a region with the lowered activation barrier for the primary endothermic reaction at different values of the parameter $p $ are shown in the Fig. 5. The curves presented in the left panel are plotted assuming $ T_a = 450 K, $ and those in the right panel result from the computations carried out at lower ambient temperature $ T_a = 300 K. $ One can observe that even at rather low exothermicity of the secondary reaction $ (\alpha = 2) $ the chemistry may develop at sufficient value of the parameter $ p. $ When $ T_a $ is close to $ T_0, $ so that the effect of the heat transfer is not very significant, the temperature behavior resembles the one appearing within the intermediate (between slow and explosive) regime for chemical reactions at $ p = 0.2, $ and moderate temperature rise is noticeable at $ p = 0.1. $ At lower $ T_a $ the effect of heat transfer must be balanced by the  effect of the external factor, and the temperature rise may occur at higher values of $ p. $ Nevertheless, the chemistry could remain self-sustained at reasonably low $ p, $ as shown in the Fig. 5 (see right panel).

\begin{figure}[t]  %%%fig. 5
\begin{center}
\includegraphics[width=8.8cm,height=4.5cm]{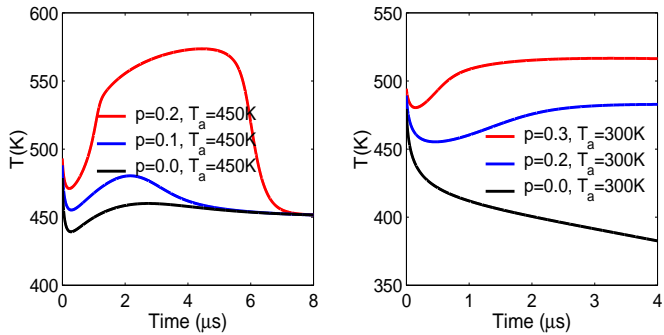}
\caption{ Temperature dependences on time in the interior of the separated region with the lowered activation barrier at different values of the parameter $ p. $ The curves are plotted  $E_1 = 1.5 eV,\ E_2 = 1eV,\  T_0 = 500K,\ \delta = 0.1,\ a = 10^4,\ \alpha = 2. $   %%  Left panel:  $ T_a = 450 K,\ p = 0.2 $ (dash-dotted line), $0.1 $ (dashed line), $ 0 $ (solid line). Right panel: $ T_a = 300 K,\ p =0.3$ (dash-dotted line), $ 0.2 $ (dashed lines) and $ 0 $ (solid lines). 
}  
\label{rateI}
\end{center}
\end{figure}

The previous analysis proves that, despite its simplicity, the adopted model enables to catch some essential features of the initiation chemistry in energetic materials. Within this model, the crucial part taken by the regions  where the activation  barrier $ E_1 $  is lowered  is elucidated using  chemical kinetics. The lowering of the activation barrier brings the increase in the first reaction rate, thus creating better opportunities for the secondary reactant accumulation. This is necessary to provide a sufficient energy release in the course of the secondary exothermic reaction, especially at moderate exothermicity of  the latter. Also, the lower is the activation energy $ E_1 $ the lower exothermicity of the secondary reaction is required to support  the chemistry. The possibilities of the local reduction of the activation barrier for the primary endothermic reaction and its importance in the initiation of the decomposition of some solid EM were theoretically explored in some  earlier works  basing on first-principles electron structure calculations. The present simple model leads to conclusions which agree with those obtained in these works. Besides, the adopted model provides us with the means to roughly compare sensitivities of energetic materials. Assuming that relevant activation energies, exothermicities and other parameters included into consideration are known for a certain EM, one may estimate the value of $ p $ required to provide self-sustaining chemistry. The greater is this value, the lower is the sensitivity of the material under consideration. In the next section we apply the described model to qualitatively analyze initial steps in the decomposition of certain practical solid explosives.

\section{iii. Application: DADNE versus TATB}

 Being in solid state, energetic nitrocompounds form layered molecular crystals. The layers are held together by rather weak van der Waals forces, whereas both intra- and intermolecular interactions within the layers are much stronger, being provided by covalent and/or hydrogen bonding. Under normal conditions these materials are chemically stable, and no spontaneous dissociation occurs. However, a significantly strong external disturbance of the system by a mechanical impact, a shock wave or a laser beam may start the chemical decomposition leading to detonation. It is commonly believed that the first step in the decomposition  of DADNE crystals is an endothermic detachment of $ NO_2 $ groups or CONO and/or HONO isomerizations \cite{36,37,38}.
   The activation energy barriers for these reactions have the order of $1.8-4.1 eV,$ which corresponds to the dissociation energy values between 40 and 90 kcal/mol. Initially, the reaction developes at discrete localized regions, which was experimentally demonstrated for some energetic molecular crystals \cite{39,40}.   
  The structures of DADNE and TATB molecular crystals were repeatedly simulated using various {\it ab initio} computational techniques \cite{31,32,41,42,43}. The results theoretically obtained for relaxed and defect-free structures demonstrate good agreement with the experiments \cite{44,45,46}. It was shown that layers in the relaxed ideal DADNE crystals are corrugated (zigzag-shaped) parallel sheets (see Fig. 6), whereas in TATB crystals the layers are nearly plane. In practical DADNE crystals the layers parallelism may be destroyed in some local regions characterized by a high concentration of dislocations, grain boundaries and other imprefections. Such regions may appear as a result of a shear-strain shifting of the adjacent layers with respect to each other along certain directions as shown in this figure. In these special regions, some molecules belonging to adjacent zigzag-chaped layers may get so close that electron orbitals for their $ NO_2 $ groups overlap. This brings noticeable changes into the electron structure within the regions \cite{31}. In DADNE crystals, the activation barrier for the endothermic $ NO_2 $ cleavage was predicted to drop from $ 4.1 eV \ (94kcal/mol) $ to $1.8 eV\ (42kcal/mol)$ as the shear-strain increases. At the same time, the CONO isometrization activation energy  was shown to remain almost independent of the strain slightly increasing from $ 1.8 eV $ in the ideal perfect crystal to $ 2.2 eV $ in the strongly deformed regions. Also, the temperature inside these regions should be higher than in the bulk of material because they are deformed to a greater extent and more heat is released there in the process of deforming \cite{32}.

Assuming that the frequency factor $ A $ takes on value of $ 10^{19} 1/s $ for the $ NO_2$ cleavage and $ 10^{12} 1/s $ for the CONO isomerization, respectively, (as given by Brill and James \cite{36}), one concludes that, in all likelihood, the primary endothermic reaction within the deformed local regions in DADNE crystals should be the $ C-NO_2 $ dissociation. Outside these deformed regions, where the crystal is comparatively free from imperfections, the CONO isomerization prevailes over the $ NO_2 $ cleavage due to the higher activation energy of the latter. So, deformed DADNE crystals may be treated as heterogeneous materials, including local regions where the first step in the decomposition chemistry is switched from the slow CONO isomerization to the fast $C - NO_2 $ bond rupture. Within these regions, the next step in the decomposition is the exothermic association $ 2NO_2 \to N_2O_4 $ with the exothermicity $ 13.8 kcal/mol $ or $18.4 kcal/mol $ depending on the configuration of the $ N_2 O_4 $ molecules \cite{29}. For this reaction, the activation energy takes on values close to $0.8 eV. $ Estimating the average exothermicity as $16.1 kcal/mol, $ we get $ \alpha = 2.2, $ which indicates low exothermicity of the secondary reaction compared to the activation energy of the primary endothermic $NO_2 $ cleavage. However, at  high temperature inside the deformed regions $ (T_0 \sim 500-750 K) $ the ratio of the reaction rates within these regions is rather high $(w_1/w_2 \sim 10^2-10^3).$ This occurs due to the slowness of the exothermic association. 

\begin{figure}[t]  %%%fig. 6
\begin{center}
\includegraphics[width=4.5cm,height=8.8cm,angle=-90]{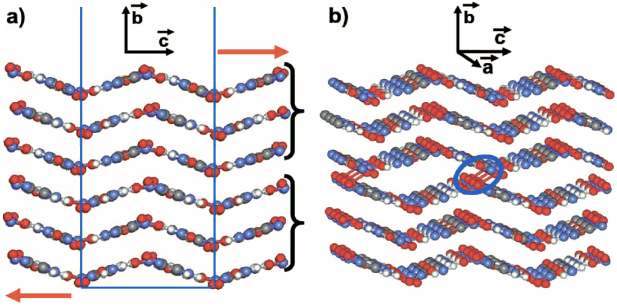}
\caption{a) Schematics of the DADNE molecular crystal cross-section by a $ \bf b  c $ plane. Vertical lines indicate the size of the periodic cell. Red arrows show the direction of the applied sheer strain shifting three lower layers with respect to three upper ones. b) Schematics of the DADNE crystal where the layers parallelism is distorted due to the strain. A local region where molecules belonging to different layers get close is indicated  (After Ref. \cite{31}).
}  
\label{rateI}
\end{center}
\end{figure}

So,  the secondary reactant may be accumulated within the deformed regions in DADNE crystals. This gives grounds to expect the chemistry initiation in such a region, provided that a reasonably moderate amount of $ C-NO_2 $ bonds are broken by the external perturbation. Computations within the suggested simple model bring results confirming these expectations. As shown in the left panel of the Fig. 7, the temperature in the deformed regions of DADNE rises when  about  15\% of the original molecules $ C-NO_2 $ bonds are ruptured by an external factor. Actually, the amount of externally dissociated molecules may be significantly smaller  because $ NO_2 $ molecules produced in the deformed regions are accumulated between the layers and their presence further lessens the activation barrier for the subsequent $ C-NO_2 $ bonds ruptures. The material apart from the special regions behaves quite differently. There, the primary endothermic reaction favors the CONO isomerization path with the low frequency factor $(A_1 = 10^{12} 1/s), $ so the ratio $ A_1/A_2 $ takes on values of the order of unity, and the reactions rates ratio $ w_1/w_2 $ is much smaller than one, which prevents the accumulation of the secondary reactants. This fact along with low exothermicity of the secondary reaction makes the material without special regions extremely low-sensitive to the effect of shear strain even when we heat a certain portion of this nearly undeformed material up to $ T_a = 500 K. $ We do not observe the temperature rise in these portions of EM even though about 50\% of the original molecules are externally dissociated (see right panel of the Fig. 7).

As for TATB molecular crystals, their electronic structure is found to be almost insensitive to the shear-strain due to near planar shape of the layers. Accordingly, the endothermic breakdown of the TATB hydrogen-bonded ring network including elimination of water molecules remains the favored pathway for the first dissociation step in the presence of the shear strain. The activation energy is estimated as  $ E_1 \approx 2.5 eV\ (60kcal/mol) $ \cite{27} regardless of the shear-strain intensity. Therefore, one cannot separate out of the bulk of TATB crystal special regions where the primary reaction switches  to $ NO_2 $ cleavage, which is accompanied by the strong acceleration of the reaction. The temperature dependences on time in the region originally heated up to $ 500 K $ reveal the features closely resembling those shown in the right panel of the Fig. 7, confirming the repeatedly reported low sensitivity of TATB, to shock-waves or impacts.

\begin{figure}[t]  %%%fig. 7
\begin{center}
\includegraphics[width=8.8cm,height=4.5cm]{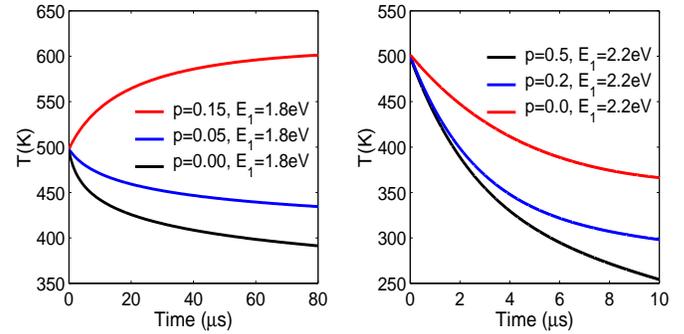}
\caption{ Temperature dependences on time in  a DADNE molecular crystal. The curves are plotted using Eqs. (\ref{8}),(\ref{10}),(\ref{16}) at $ E_2 = 0.8 eV,\ T_a = 500 K,\ \alpha = 2.2. $ Left panel: Deformed region; $T_0 = 500 K,\ T_a = 300K,\  E_1 = 1.8 eV,\ a = 10^7$. Right panel: Undeformed region; $E_1 = 2.2 eV,\ a= 1. $
}  
\label{rateI}
\end{center}
\end{figure}

We may conclude that  initial steps in the ignition of DADNE and TATB molecular crystals exposed to a strong shear-strain which may be caused by a low velocity mechanical impact or a shock wave, significantly differ. In DADNE crystals the strain generates formation of small regions characterized by high density of dislocations and other imperfections where the activation barriers for the endothermic $ C-NO_2 $ ruptures are significantly lowered. In these regions, the primary endothermic reaction is strongly accelerated, which creates favorable conditions for the accumulation of $ NO_2 $ molecules. The following association of these molecules is accompanied by the energy release which may be sufficient to support the decomposition chemistry provided that a certain moderately small portion of the original molecules in these regions is dissociated by the external perturbation. This gives grounds to conjecture that the above regions may serve as hot spots for the DADNE ignition for the considered ignition mechnism.	

At the same time, stable molecular and crystalline structure of TATB prevents formation of regions where the acceleration of the primary endothermic reaction occurs, assuming the shear-strain magnitudes sufficient for DADNE. Possibly, the desired result may be achieved at higher strain magnitudes. One may put forward this conjecture basing on the experimentally established fact, which implies that the threshold pressure of the detonation initiation in TATB is approximately ten times higher than that typical for other high explosive EMs \cite{47}. %%From the microscopical point of view, one may suggest that the desirable changes in the local electronic structure of TATB could appear as an effect of large amount of interstitial $ NO_2 $ molecules distorting planar shape of layers \cite{16}. These $ NO_2 $ molecules must be created by the external perturbation rupturing $ C-NO_2 $ bonds. Only when a significant part of the original molecules is consumed in such a way, the switching of the primary reaction to the fast $ C-NO_2 $ cleavage may happen in some regions, and the exothermic chemistry could start there.

\section{iv. Conclusion}

In conclusion, progress in understanding of the nature of hot spots in  solid energetic materials is difficult owing to the extreme complexity of both physical mechanisms and chemistry involved in the ignition process. Here, we developed our analysis basing on the earlier suggestion that in some energetic molecular crystals such as DADNE, hot spots for the low velocity impact ignition mechanisms may be small regions where the activation energy barriers for the primary endothermic reaction are significantly reduced. In DADNE crystals the barriers lowering may originate from the local bindings of molecules belonging to the adjacent corrugated layers of the molecular crystal, appearing under high shear strain applied along certain directions within the layers planes. We show that the difference in the activation energies within and without such a region may be crucial for the secondary reactant accumulation thus providing favorable conditions for the decomposition chemistry to develop within a hot spot region while it is still not possible in the bulk of material. We concentrate on the case when the external perturbation is short-lived, so the consumption of a major part of the primary reactant is sustained by the energy released in the course of first exothermic reactions closely following the primary endothermic step. We show that when the endothermic reaction rate significantly exceeds the rate of the subsequent exothermic reaction, the chemistry may develop within a hot spot region even in the case of moderate/low exothermicity of the secondary reaction $ (\alpha > 1).$ Our model given by Eqs.(\ref{7})--(\ref{9}) essentially simplifies the actual  chemistry. However, the model may be  generalized to include more chemical reactions which take important parts in the ignition chemistry of a particular energetic molecular crystal. Also, the energy transfer between the hot spot region and its environment and within the hot spot region itself may be approximated more thoroughly. Nevertheless, we do believe that the simple model used in the present work enables us to catch some important and essential features of the early stage of the detonation initiation in energetic molecular crystals. The model is applied to analyze comparative sensitivity of DADNE and TATB molecular crystals.
\vspace{4mm}

{\bf Acknowledgments:} 
Author  thanks M. Kuklja for very helpful discussion and  G. M. Zimbovsky for help with the manuscript preparation.  This work was partly supported  by  NSF-DMR-PREM  1523463.

%\begin{widetext}  \end{widetext}

\end{document}